\documentclass[twocolumn,showpacs,preprintnumbers,amssymb,nofootinbib]{revtex4}

\usepackage{graphicx}% Include figure files
\usepackage{bm} % Include bold math: \bm{} creates bold letters and symbols in math mode

\begin{document}

%\preprint{DF/IST-04.2002}

\title{Nariai,
Bertotti-Robinson and anti-Nariai solutions in higher dimensions}

\author{Vitor Cardoso}
\email{vcardoso@fisica.ist.utl.pt} \affiliation{ Centro
Multidisciplinar de Astrof\'{\i}sica - CENTRA, Departamento de
F\'{\i}sica, Instituto Superior T\'ecnico, Av. Rovisco Pais 1,
1049-001 Lisbon, Portugal \\ \&\\ Centro de F\'{\i}sica
Computacional, Universidade de Coimbra, P-3004-516 Coimbra,
Portugal}
\author{\'Oscar J. C. Dias}
\email{oscar@fisica.ist.utl.pt} \affiliation{ Centro
Multidisciplinar de Astrof\'{\i}sica - CENTRA, Departamento de
F\'{\i}sica, Instituto Superior T\'ecnico, Av. Rovisco Pais 1,
1049-001 Lisbon, Portugal \\ \&\\ CENTRA, Departamento de
F\'{\i}sica, F.C.T., Universidade do Algarve, Campus de Gambelas,
8005-139 Faro, Portugal}
\author{Jos\'e P. S. Lemos}
\email{lemos@kelvin.ist.utl.pt} \affiliation{ Centro
Multidisciplinar de Astrof\'{\i}sica - CENTRA, Departamento de
F\'{\i}sica, Instituto Superior T\'ecnico, Av. Rovisco Pais 1,
1049-001 Lisbon, Portugal}

\date{\today}
\begin{abstract}
We find all the higher dimensional solutions of the
Einstein-Maxwell theory that are the topological product of two
manifolds of constant curvature. These solutions include the
higher dimensional Nariai, Bertotti-Robinson and anti-Nariai
solutions, and the anti-de Sitter Bertotti-Robinson solutions
 with toroidal and hyperbolic
topology (Pleba\'nski-Hacyan solutions). We give explicit results
for any dimension $D\geq 4$. These solutions are generated from
the appropriate extremal limits of the higher dimensional
near-extreme black holes in a de Sitter, and anti-de Sitter
backgrounds. Thus, we also find the mass and the charge parameters
of the higher dimensional extreme black holes as a function of the
radius of the degenerate horizon.
\end{abstract}

\pacs{04.20.Jb, 04.70.Bw, 04.20.Gz}


\maketitle

%%%%%%%%%%%%%%%%%%%%%%%%%%%%%%%%%%%%%%%%%%%%%%%%%%%%%%%%%%%%%%%%%%%%%%%%%%%%%%%%%%%%%%%
\section{\label{sec:Int}Introduction}
%%%%%%%%%%%%%%%%%%%%%%%%%%%%%%%%%%%%%%%%%%%%%%%%%%%%%%%%%%%%%%%%%%%%%%%%%%%%%%%%%%%%%%%

The interest on higher dimensional spacetimes was boosted with the
development of string theories. More recently, there has been a
renewed interest in connection to the TeV-scale theory
\cite{hamed} which suggests that the universe in which we live may
have large extra dimensions. According to this conjecture, we
would live on a four-dimensional sub-manifold, where the Standard
Model inhabits, whereas the gravitational degrees of freedom
propagate throughout all dimensions. This has motivated a wide
search for various phenomena \cite{tails} involving higher
dimensions. In particular, it is possible that future
accelerators, such as the Large Hadron Collider (LHC) at CERN
produce black holes, and thus detect indirectly gravitational
waves \cite{bhprod}.

In this paper we deal with exact solutions of the Einstein-Maxwell
theory in higher dimensions. The higher dimensional counterparts
of the Schwarzschild and of the Reissner-Nordstr\"{o}m black holes
$-$ the Tangherlini black holes $-$ have been found and discussed
in \cite{tangherlini}. The $D$-dimensional Majumdar-Papapetrou
black holes have been found in \cite{myersMP} (see also
\cite{lemoszanchinMP}). The higher dimensional Kerr black hole $-$
the Myers-Perry black hole $-$ was found in \cite{myersperry} and
further discussed in \cite{EmparanMyers}. The higher dimensional
counterpart of the Kerr-Newman black hole is not yet known (see
\cite{IdaUchida} for a discussion). The higher dimensional
Schwarzschild and Reissner-Nordstr\"{o}m black holes in an
asymptotically de Sitter (dS) spacetime and in an asymptotically
anti-de Sitter (AdS) spacetime have also been discussed in
\cite{tangherlini}. Now, in an asymptotically AdS 4-dimensional
background, besides the black holes with spherical topology, there
are also solutions with planar, cylindrical or toroidal topology
found and discussed in \cite{Toroidal_4D} (neutral case),
\cite{Toroidal_elect_4D} (electric case), and
\cite{Toroidal_mag_4D} (magnetic case), and black holes with
hyperbolic topology analyzed in \cite{topological}. The higher
dimensional extensions of these non-spherical AdS black holes are
already known. Namely, the $D$-dimensional AdS black holes with
planar, cylindrical or toroidal topology were discussed in
\cite{Birmingham-TopDdim} (neutral case), in
\cite{Toroidal_Q_Ddim} (electric case) and in
\cite{Dehghani-Toroidal_magJ_Ddim} (magnetic case), and the
$D$-dimensional AdS black holes with hyperbolic topology were
analyzed in  \cite{Birmingham-TopDdim,DehghaniRotTopol}.

In a 4-dimensional background with generic cosmological constant,
and still in the context of Einstein-Maxwell theory, there are
other interesting solutions that do not contain a black hole, but
are the direct topological product of two manifolds of constant
curvature. These are the Nariai solution \cite{Nariai}, the
Bertotti-Robinson solution \cite{BertRob}, the anti-Nariai
solution \cite{CaldVanZer}, and the Pleba\'nski-Hacyan solutions
\cite{PlebHac}.  For a detailed historical overview on these
solutions and for references see, e.g.,
\cite{Bousso60y,OrtaggioPodolsky,OscLem_nariai}. The discussion of
these solutions in a more mathematical context can be found in
\cite{FickenCahenDefrise}. Some of these solutions, but not all,
have already been discussed in a higher dimensional spacetime.

Ginsparg and Perry \cite{GinsPerry} (see also \cite{Nariai_Q})
have connected the extreme dS-Schwarzschild black hole with the
Nariai solution \cite{Nariai} in a 4-dimensional spacetime. That
is, they have shown that the already known Nariai solution (which
is not a black hole solution) could be generated from an
appropriate extremal limit of a near-Nariai black hole. They
realized this connection while they were studying the quantum
stability of the Nariai and the dS-Schwarzschild solutions. A
similar procedure allows to generate the Bertotti-Robinson, the
anti-Nariai, and the Pleba\'nski-Hacyan solutions from appropriate
near-extreme black holes.

In this paper we shall use the procedure introduced in
\cite{GinsPerry} in order to generate, from the appropriate
extremal limits of the $D$-dimensional near-extreme black holes,
all the higher dimensional solutions that are the topological
product of two manifolds of constant curvature. These solutions
include the higher dimensional Nariai, Bertotti-Robinson,
anti-Nariai, and Pleba\'nski-Hacyan solutions. We give explicit
results for any dimension $D\geq 4$. In passing we also find the
values of the mass and of the charge for which one has extreme
black holes in an asymptotically dS and in an asymptotically AdS
higher dimensional spacetime. This analysis has already been
initiated in \cite{KodamaIshibashi}, but our procedure and results
are complementary to those of \cite{KodamaIshibashi}. In the AdS
case, the discussion carried in this paper includes black holes
with spherical topology, with planar, cylindrical and toroidal
topology, and with hyperbolic topology.

The plan of this paper is as follows. In Sec. \ref{sec:Action}, we
set the Einstein-Maxwell action in a $D$-dimensional background.
In Sec. \ref{sec:Solutions D-dim cosmolog dS} we discuss the
properties of the extreme higher dimensional dS black holes, and
we find their extremal limits, i.e., the associated Nariai-like
solutions. In Sec. \ref{sec:Solutions D-dim cosmolog AdS} we do
the same but this time in an AdS background.

%%%%%%%%%%%%%%%%%%%%%%%%%%%%%%%%%%%%%%%%%%%%%%%%%%%%%%%%%%
%%%%%%%%%%%%%%%%%%%%%%%%%%%%%
\section{\label{sec:Action}Action and equations of motion}
%%%%%%%%%%%%%%%%%%%%%%%%%%%%%
%%%%%%%%%%%%%%%%%%%%%%%%%%%%%%%%%%%%%%%%%%%%%%%%%%%%%%%%%%

We will discuss solutions that are the extremal limits of the
near-extreme cases of the static higher dimensional black holes.
Some of these black holes were found by Tangherlini
\cite{tangherlini}, and are the higher dimensional cousins of the
Schwarzschild and of the Reissner-Nordstr\"{o}m black holes. We
work in the context of the Einstein-Maxwell action with a
cosmological constant $\Lambda$,
\begin{eqnarray}
I=\frac{1}{16\pi}\! \int_{\cal M} \!\! d^Dx\sqrt{-g} \left (
R-\frac{(D-1)(D-2)}{3}\,\Lambda -F^2 \right ), \nonumber \\
 \label{I D-dim}
\end{eqnarray}
where $D$ is the dimension of the spacetime, $g$ is the
determinant of the metric $g_{\mu\nu}$, $R$ is the Ricci scalar,
and $F_{\mu\nu}=\partial_{\mu}A_{\nu}-\partial_{\nu}A_{\mu}$ is
the Maxwell field strength of the gauge field $A_{\nu}$.
 We set the $D$-dimensional Newton's constant equal to
one, and $c=1$. The variation of (\ref{I D-dim}) yields the
equations for the gravitational field and for the Maxwell field,
respectively,
\begin{eqnarray}
& &\hspace{-1cm}
R_{\mu\nu}-\frac{1}{2}R\,g_{\mu\nu}+\frac{(D-1)(D-2)}{6} \,\Lambda
g_{\mu\nu}=8\pi\,T_{\mu\nu},
\nonumber \\
& & \hspace{-1cm}
 \nabla_{\mu} F^{\mu\nu}=0,
 \label{D-dim:eqs of motion Einstein}
\end{eqnarray}
where $R_{\mu\nu}$ is the Ricci tensor and $T_{\mu\nu}$ is the
electromagnetic energy-momentum tensor,
\begin{eqnarray}
T_{\mu\nu}=\frac{1}{4\pi}\left ( g^{\alpha \beta} F_{\alpha \mu}
F_{\beta \nu}-\frac{1}{4}g_{\mu\nu}F_{\alpha\beta}F^{\alpha\beta}
\right ) \:.
 \label{D-dim:energy tensor}
\end{eqnarray}
In (\ref{I D-dim}), the coefficient of $\Lambda$ was chosen in
order to insure that, for any dimension $D$, the pure dS or pure
AdS spacetimes are described by $g_{tt}=1-(\Lambda/3)r^2$, as
occurs with $D=4$.

%The contraction of the first relation of (\ref{D-dim:eqs of motion Einstein}) with
%$g^{\mu\nu}$ yields for the Ricci tensor
%\begin{eqnarray}
%R=\frac{D(D-1)}{3}\,\Lambda-\frac{16\pi}{D-2}\,T \:,
% \label{D-dim:Ricci Tensor}
%\end{eqnarray}
%where $T$ is the trace of $T_{\mu\nu}$.
% We remark that in a general $D$-dimensional background the electromagnetic
%energy-momentum tensor is not traceless. Indeed, the contraction
%of (\ref{D-dim:energy tensor}) with $g^{\mu\nu}$ yields
%\begin{eqnarray}
%T=-\frac{D-4}{4\pi}\,F_{\mu\nu}F^{\mu\nu} \:,
% \label{D-dim:trace T}
%\end{eqnarray}
%which vanishes only for $D=4$.

%%%%%%%%%%%%%%%%%%%%%%%%%%%%%%%%%%%%%%%%%%%%%%%%%
\section{\label{sec:Solutions D-dim cosmolog dS}Higher dimensional
extreme \lowercase{d}S black holes and Nariai-like solutions}
%%%%%%%%%%%%%%%%%%%%%%%%%%%%%%%%%%%%%%%%%%%%%%%%%

In order to generate the higher dimensional Nariai,
dS$-$Bertotti-Robinson,  and Nariai$-$Bertotti-Robinson solutions
one needs first to carefully find the values of the mass and of
the charge for which one has an extreme dS black hole. We will do
this in Sec. \ref{sec:BH D-dim dS}, and in Sec. \ref{sec:BH D-dim
extremal limits dS} we will generate the Nariai-like solutions
from the extremal limits of the near-extreme black holes.

%%%%%%%%%%%%%%%%%%%%%%%%%%%%%%%%%%%%%%%%%%%%%%%%%
\subsection{\label{sec:BH D-dim dS}Higher dimensional
extreme black holes in an asymptotically \lowercase{d}S
background}
%%%%%%%%%%%%%%%%%%%%%%%%%%%%%%%%%%%%%%%%%%%%%%%%%

In an asymptotically de Sitter background, $\Lambda> 0$, the most
general static higher dimensional black hole solution with
spherical topology was found by Tangherlini \cite{tangherlini}.
The gravitational field is
\begin{equation}
ds^{2}=-f(r)dt^{2}+f(r)^{-1}dr^{2}+r^{2}\,d\Omega_{D-2}^2
 \label{RN:metric D-dim}
\end{equation}
where $d\Omega_{D-2}^2$ is the line element on an unit
$(D-2)$-sphere,
\begin{equation}
d\Omega^2_{D-2}=d\theta_1^2+\sin^2\theta_1\,d\theta_2^2+ \cdots
+\prod_{i=1}^{D-3}\sin^2\theta_i\,d\theta_{D-2}^2\:,
 \label{metric sphere D-dim}
\end{equation}
and the function $f(r)$ is given by
\begin{equation}
f(r)=1-\frac{\Lambda}{3}r^2-\frac{M}{r^{D-3}}+\frac{Q^2}{r^{2(D-3)}}\:,
 \label{RN:f cosmolog D-dim}
\end{equation}
The mass parameter $M$ and the charge parameter $Q$ are related to
the ADM mass, $M_{\rm ADM}$, and ADM electric charge, $Q_{\rm
ADM}$, of the solution by \cite{myersperry}
\begin{eqnarray}
& &M_{\rm ADM}=\frac{(D-2)\Omega_{D-2}}{16\pi}\,M \:,\nonumber \\
& &Q_{\rm ADM}=\sqrt{\frac{(D-3)(D-2)}{2}}\,Q \:,
 \label{ADM hairs D-dim}
\end{eqnarray}
where $\Omega_{D-2}$ is the area of an unit $(D-2)$-sphere,
\begin{equation}
\Omega_{D-2}=\frac{2\pi^{(D-1)/2}}{\Gamma[(D-1)/2]}\:.
\label{integratedsolidangle}
\end{equation}
Here, $\Gamma[z]$ is the gamma function. For our purposes we need
to know that $\Gamma[z]=(z-1)!$ when $z$ is a positive integer,
$\Gamma[1/2]=\sqrt{\pi}$, and $\Gamma[z+1]=z\Gamma[z]$. The radial
electromagnetic field produced by the electric charge $Q_{\rm
ADM}$ is given by
\begin{equation}
F=-\frac{Q_{\rm ADM}}{r^{D-2}}\,dt\wedge dr\:.
 \label{RN:maxwell D-dim}
\end{equation}
These solutions have a curvature singularity at the origin, and
the black hole solutions can have at most three horizons, the
Cauchy horizon $r_-$, the event horizon $r_+$ and the cosmological
horizon $r_{\rm c}$, that satisfy $r_-\leq r_+ \leq r_{\rm c}$.

We are now interested in the $D$-dimensional extreme
dS-Tangherlini black holes. That is, in order to start searching
for Nariai-like solutions one needs first to carefully find the
parameters $M$ and $Q$ as a function of the degenerate horizon. To
settle the nomenclature and the technical procedure, we start with
the five-dimensional case, $D=5$. We look to the extreme dS black
holes, for which two of the horizons coincide. Let us label this
degenerate horizon by $\rho$. In this case, and for $D=5$, the
function $f(r)$ given by (\ref{RN:f cosmolog D-dim}) can be
written as
\begin{eqnarray}
f(r)=-\frac{\Lambda}{3}\frac{1}{r^4}(r-\rho)^2(r+\rho)^2
 \left ( r^2-\frac{3}{\Lambda}+2\rho^2\right ).
 \label{Fextreme 5D}
 \end{eqnarray}
Thus, besides the degenerate horizon $r=\rho$, there is another
horizon at $\sigma =\sqrt{3/\Lambda-2\rho^2}$.
%\begin{eqnarray}
%\sigma =\sqrt{\frac{3}{\Lambda}-2\rho^2} \:.
%  \label{zero extra 5D}
% \end{eqnarray}
From (\ref{RN:f cosmolog D-dim}) with $D=5$ and
 (\ref{Fextreme 5D}), the mass parameter $M$ and the charge parameter $Q$ of the
black holes can be written as functions of $\rho$: $M=\rho^2
(2-\Lambda \rho^2)$ and $Q^2 =\rho^4 \left (1- 2\Lambda\rho^2/3
\right )$.
%\begin{eqnarray}
%M=\rho^2 (2-\Lambda \rho^2)\:, \:\:\: {\rm and} \:\:\:
% Q^2 =\rho^4 \left (1- \frac{2\Lambda}{3}\rho^2 \right )\:.
% \label{mq 5D}
% \end{eqnarray}
The condition $Q^2\geq 0$ implies that $\rho \leq
\sqrt{3/(2\Lambda)}$. At this point we note that $M$ and $Q$ first
increase with $\rho$ (this sector corresponds to $\sigma>\rho$),
until $\rho$ reaches the critical value $\rho=\sqrt{1/\Lambda}$
(this sector corresponds to $\sigma=\rho$), and then $M$ and $Q$
start decreasing until $\rho$ reaches its maximum allowed value
(this sector corresponds to $\sigma<\rho$). These three sectors
are associated to three distinct extreme dS black holes: the cold,
the ultracold and the Nariai black holes, respectively (here we
follow the nomenclature used in the analogous 4-dimensional black
holes \cite{Rom}. Note that the Nariai black hole differs from the
Nariai solution which is not a black hole solution). More
precisely, for $0<\rho<1/\sqrt{\Lambda}$ one has the cold black
hole with $r_-=r_+ \equiv \rho$ and $r_{\rm c}\equiv \sigma$. The
ranges of the mass and charge parameters for the cold black hole
are $0<M<1/\Lambda$ and $0<Q<1/(\sqrt{3}\,\Lambda)$. The case
$\rho=1/\sqrt{\Lambda}$ gives the ultracold black hole in which
the three horizons coincide, $r_-=r_+ = r_{\rm c}$. Its mass and
charge parameters are $M=1/\Lambda$ and $Q=1/(\sqrt{3}\,\Lambda)$.
For $1/\sqrt{\Lambda}<\rho\leq \sqrt{3/(2\Lambda)}$ one has the
Nariai black hole with $r_+=r_{\rm c} \equiv \rho$ and $r_-\equiv
\sigma$. The ranges of the mass and charge parameters for the
Nariai black hole are $3/(4\Lambda)\leq M<1/\Lambda$ and $0\leq
Q<1/(\sqrt{3}\,\Lambda)$.

Now, the above construction can be extended to $D$-dimensional
extreme dS black holes. In the extreme case the function $f(r)$
given by (\ref{RN:f cosmolog D-dim}) can be written as
\begin{eqnarray}
f(r)= (r-\rho)^2 \frac{1}{r^2}\left [ 1- \frac{\Lambda}{3}
 \left [ r^2+h(r) \right ] \right ]\:,
 \label{Fextreme D-dim}
 \end{eqnarray}
where $r=\rho$ is the degenerate horizon of the black hole, and
\begin{eqnarray}
h(r)=a+br+\frac{c_1}{r}+\frac{c_2}{r^2}+\cdots+\frac{c_{2(D-4)}}{r^{2(D-4)}}
\:,
  \label{aux function h(r)}
 \end{eqnarray}
where $a, b, c_1,...,c_{2(D-4)}$ are constants that can be found
through the matching between (\ref{RN:f cosmolog D-dim}) and
(\ref{aux function h(r)}). This procedure yields the mass
parameter $M$ and the charge parameter $Q$ of the black holes as a
function of $\rho$,
\begin{eqnarray}
& &M=2\rho^{D-3} \left ( 1-\frac{D-2}{D-3}\,\frac{\Lambda}{3}
\rho^2
\right)\:, \nonumber \\
& & Q^2 =\rho^{2(D-3)} \left (
1-\frac{D-1}{D-3}\,\frac{\Lambda}{3} \rho^2 \right)\:.
 \label{mq D-dim}
 \end{eqnarray}
The condition $Q^2\geq 0$ implies that $\rho \leq \rho_{\rm max}$
with
\begin{eqnarray}
 \rho_{\rm max}=\sqrt{\frac{D-3}{D-1}\,\frac{3}{\Lambda}}\:.
 \label{def rho max}
\end{eqnarray}

\begin{figure}[ht]
\includegraphics*[height=5cm]{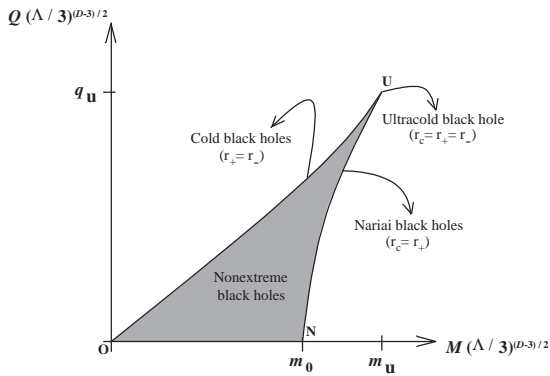}
   \caption{\label{range mq dS bh D-dim}
Range of $M$ and $Q$ for which one has a nonextreme black hole
(region interior to the closed line $ONUO$), an extreme Nariai
black hole with $r_+ = r_{\rm c}$ (line $NU$), an extreme cold
black hole with $r_- =r_+$ (line $OU$), and an extreme ultracold
black hole with $r_-=r_+ = r_{\rm c}$ (point $U$). The line $ON$
represents the nonextreme dS-Schwarzschild black hole, and point
$N$ represents the extreme Nariai Schwarzschild black hole. The
non-shaded area represents a naked singularity region. The
constants in the axes are $m_0=\frac{2}{D-1}\left (
\frac{D-3}{D-1}\right )^{(D-3)/2}$, $m_{\rm u}=\frac{4}{D-1}\left
( \frac{(D-3)^2}{(D-2)(D-1)}\right )^{(D-3)/2}$, and
 $q_{\rm u}=\frac{1}{\sqrt{D-2}}\left ( \frac{(D-3)^2}{(D-2)(D-1)}
  \right )^{(D-3)/2}$.
 }
\end{figure}

For the $D$-dimensional cold black hole ($r_-=r_+$), $M$ and $Q$
increase with $\rho$, and one has
\begin{eqnarray}
 & & 0<\rho<\rho_{\rm u}\,, \qquad
 0<M<\frac{4}{D-1}\,  \rho_{\rm u}^{\, D-3}\,, \nonumber \\
 & & 0<Q<\frac{1}{\sqrt{D-2}}\, \rho_{\rm u}^{\, D-3}\:,
 \label{ColdDdim:range}
\end{eqnarray}
where we have defined
\begin{eqnarray}
 \rho_{\rm u}=\sqrt{\frac{3}{\Lambda}}\,\frac{D-3}{\sqrt{(D-2)(D-1)}}\:.
 \label{def rho ultra}
\end{eqnarray}
For the $D$-dimensional ultracold black hole
 ($r_-=r_+=r_{\rm c}$), one has
 \begin{eqnarray}
 & & \rho=\rho_{\rm u}\,, \qquad M=\frac{4}{D-1} \,  \rho_{\rm u}^{\,D-3}\,,
 \nonumber \\
 & & Q=\frac{1}{\sqrt{D-2}}\,  \rho_{\rm u}^{\,D-3}\:.
 \label{UltracoldDdim:range}
\end{eqnarray}
Finally, for the $D$-dimensional Nariai black hole
 ($r_+=r_{\rm c}$), $M$ and $Q$ decrease with $\rho$, and one has
\begin{eqnarray}
 & & \rho_{\rm u}< \rho \leq \rho_{\rm max}\,, \quad
 \frac{2}{D-1}\,\rho_{\rm max}^{D-3}\leq M<\frac{4}{D-1} \, \rho_{\rm u}^{\,D-3}\,,
\nonumber \\
 & &  0\leq Q<\frac{1}{\sqrt{D-2}} \, \rho_{\rm u}^{\,D-3}\:.
 \label{NariaiDdim:range}
\end{eqnarray}
The ranges of $M$ and $Q$ that represent each one of the above
extreme black holes is sketched in Fig.
 \ref{range mq dS bh D-dim}. This figure and the associated
relations (\ref{mq D-dim})-(\ref{NariaiDdim:range}) are not the
main results of this paper. However, they constitute new results
that we had to find in our way into the generation of the
Nariai-like solutions. For an alternative and complementary
description of the extreme dS-Tangherlini black holes see
\cite{KodamaIshibashi}.

\begin{figure}[h]
\includegraphics*[height=14cm]{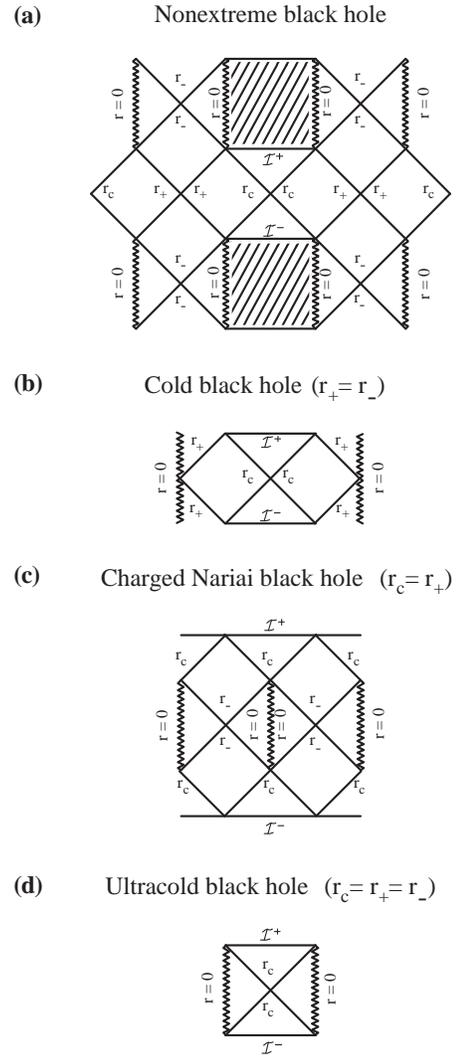}
   \caption{\label{Fig spheric Q dS}
Carter-Penrose diagrams of the dS$-$Reissner-Nordstr\"{o}m ($Q\neq
0$) black holes. The zigzag line represents a curvature
singularity, ${\cal I}$ represents the infinity ($r=\infty$),
$r_{\rm c}$ represents a cosmological horizon, $r_+$ represents a
black hole event horizon, and $r_-$ represents a Cauchy horizon.
Fig. (a) was presented in \cite{GibHawdSbh}. As far as we know,
Figs. (b)-(d) are first shown here.
 }
\end{figure}

\begin{figure}[h]
\includegraphics*[height=4cm]{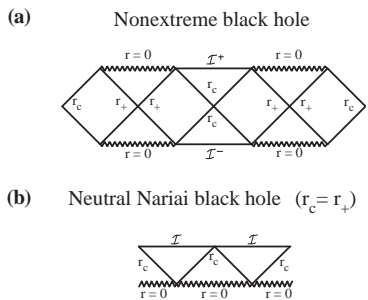}
   \caption{\label{Fig spheric dS}
Carter-Penrose diagrams of the dS-Schwarzschild ($Q=0$) black
holes. The zigzag line represents a curvature singularity, ${\cal
I}$ represents the infinity ($r=\infty$), $r_{\rm c}$ represents a
cosmological horizon, and $r_+$ represents a black hole event
horizon. Fig. (a) was presented in \cite{GibHawdSbh}, and Fig. (b)
was presented in \cite{Lake}.
 }
\end{figure}

The Carter-Penrose diagrams of the $D$-dimensional dS-Tangherlini
black holes are similar to the ones of their 4-dimensional
counterparts and are sketched in Fig. \ref{Fig spheric Q dS} in
the charged case, and in Fig. \ref{Fig spheric dS} in the neutral
case. In these diagrams each point represents a ($D-2$)-sphere of
radius $r$.


%%%%%%%%%%%%%%%%%%%%%%%%%%%%%%%%%%%%%%%%%%%%%%%%%
\subsection{\label{sec:BH D-dim extremal limits dS}Extremal
 limits of the higher dimensional \lowercase{d}S black holes}
%%%%%%%%%%%%%%%%%%%%%%%%%%%%%%%%%%%%%%%%%%%%%%%%%

In this section, we apply the near extremal procedure of Ginsparg
and Perry \cite{GinsPerry} to the extreme black holes discussed in
the last subsection, in order to find the higher dimensional
Nariai, dS$-$Bertotti-Robinson, $\Lambda=0$ Bertotti-Robinson and
Nariai$-$Bertotti-Robinson solutions. The higher dimensional
Nariai solution has already been found in
\cite{CaldVanZer,KodamaIshibashi}. Here we show that it can be
generated from the near-Nariai black hole following the procedure
of \cite{GinsPerry}. Therefore, we give a new emphasis to the
solution and we set the nomenclature for the other new cases.

%%%%%%%%%%%%%%%%%%%%%%%%%%%%%%%%%%%%%%%%%%%%%%%%%
\subsubsection{\label{sec:BH D-dim Nariai}Higher dimensional Nariai
solution}
%%%%%%%%%%%%%%%%%%%%%%%%%%%%%%%%%%%%%%%%%%%%%%%%%

In order to generate the higher dimensional Nariai solution from
the near-Nariai black hole we first go back to
 (\ref{Fextreme D-dim}) and rewrite it in the form $f(r)=-A(r)(r-\rho)^2$,
where $r=\rho$ is the degenerate horizon of the black hole, and
$A(r)$ is a polynomial function of $r$. Then, we  set
$r_+=\rho-\varepsilon$ and $r_{\rm c}=\rho+\varepsilon$, where
$\varepsilon<<1$ measures the deviation from degeneracy, and the
limit $r_+\rightarrow r_{\rm c}$ is obtained when $\varepsilon
\rightarrow 0$. Now, we introduce a new time coordinate $T$, $t=
T/(\varepsilon A)$, and a new radial coordinate $\chi$,
$r=\rho+\varepsilon \cos\chi$, where $\chi=0$ and $\chi=\pi$
correspond, respectively, to the horizons $r_{\rm c}$ and $r_+$,
and $A\equiv A(\rho)=\rho^{-2}\left [ 1-
 \left ( \rho^2+h \right )\Lambda/3 \right ]>0$, with
$h\equiv h(\rho)$ defined in (\ref{aux function h(r)}). Then, in
the limit $\varepsilon \rightarrow 0$, from  (\ref{RN:metric
D-dim}) and (\ref{Fextreme D-dim}), we obtain the gravitational
field of the Nariai solution
\begin{eqnarray}
d s^2 = \frac{1}{A} \left (-\sin^2\chi\, dT^2 +d\chi^2 \right ) +
 \frac{1}{B}\, d\Omega_{D-2}^2 \:,
 \label{D-dim:Nariai solution}
\end{eqnarray}
where $\chi$ runs from $0$ to $\pi$, and $A$ and $B=1/\rho^2$ are
related to $\Lambda$ and $Q$ by
\begin{eqnarray}
& & \Lambda = \frac{3}{(D-2)(D-1)}\left [ A+(D-3)^2 B \right ]\:,
\nonumber \\
& & Q^2=\frac{(D-3)B-A}{(D-3)(D-2)B^{D-2}}\:.
 \label{D-dim:Nariai solution:range AB}
\end{eqnarray}
The Maxwell field (\ref{RN:maxwell D-dim}) of the higher
dimensional Nariai solution is
 \begin{eqnarray}
 F=Q_{\rm ADM}\,\frac{B^{(D-2)/2}}{A}\,\sin \chi \,dT \wedge d\chi\:.
\label{D-dim:Nariai Maxwell}
\end{eqnarray}
So, if we give the parameters $\Lambda$, and $Q$ we can construct
the higher dimensional Nariai solution, which is an exact solution
of Einstein-Maxwell equations (\ref{D-dim:eqs of motion Einstein})
with $\Lambda>0$ in $D$-dimensions. Through a redefinition of
coordinates, $\sin^2\chi=1-A\,R^2$ and $\tau=\sqrt{A} \, T$, the
spacetime (\ref{D-dim:Nariai solution}) can be rewritten in new
static coordinates as
\begin{eqnarray}
d s^2 = - (1-A\,R^2)\, dT^2 +\frac{dR^2}{1-A\,R^2} + \frac{1}{B}\,
d\Omega_{D-2}^2 \,,
 \label{D-dim:qNariai2}
\end{eqnarray}
and the electromagnetic field changes also accordingly to the
coordinate transformation. Written in these coordinates, we
clearly see that the Nariai solution is the direct topological
product of $dS_2 \times S^{D-2}$, i.e., of a (1+1)-dimensional dS
spacetime with a ($D-2$)-sphere of fixed radius $B^{-1/2}$. This
spacetime is homogeneous with the same causal structure as
(1+1)-dimensional  dS spacetime, but it is not an asymptotically
4-dimensional dS spacetime since the radius of the ($D-2$)-sphere
is constant ($B^{-1/2}$), contrarily to what happens in the dS
solution where this radius increases as one approaches infinity.

The neutral Nariai solution ($Q=0$) satisfies the relations
$A=\Lambda(D-1)/3$ and $B=\Lambda(D-1)/(3D-9)$. The $\Lambda=0$
limit of the Nariai solution is $D$-dimensional Minkowski
spacetime as occurs with the $D=4$ solution (see
\cite{OscLem_nariai}).

\begin{figure}[h]
\includegraphics*[height=2cm]{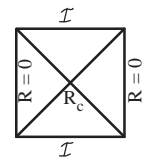}
   \caption{\label{Fig Ddim Nariai}
Carter-Penrose diagram of the Nariai solution (charged or
neutral). The zigzag line represents a curvature singularity,
${\cal I}$ represents the infinity ($R=\infty$), $R_{\rm c}$
represents a cosmological horizon.
 }
\end{figure}

The Carter-Penrose diagram of the $D$-dimensional Nariai solution
(charged or neutral) is sketched in Fig. \ref{Fig Ddim Nariai}. In
this diagram any point represents a ($D-2$)-sphere with fixed
radius $B^{-1/2}$. For a construction that starts with the
Carter-Penrose diagram of the dS black hole and leads to the
diagram of the Nariai solution see \cite{OscLem_nariai}.

%%%%%%%%%%%%%%%%%%%%%%%%%%%%%%%%%%%%%%%%%%%%%%%%%
\subsubsection{\label{sec:BH D-dim dS Bertotti-Robinson}Higher
dimensional dS Bertotti-Robinson solution}
%%%%%%%%%%%%%%%%%%%%%%%%%%%%%%%%%%%%%%%%%%%%%%%%%

In order to generate the higher dimensional dS Bertotti-Robinson
solution from the near-cold black hole we first go back to
(\ref{Fextreme D-dim}) and rewrite it in the form
$f(r)=A(r)(r-\rho)^2$, where $r=\rho$ is the degenerate horizon of
the black hole, and $A(r)$ is a polynomial function of $r$. Then,
we set $r_-=\rho-\varepsilon$ and $r_+=\rho+\varepsilon$, where
$\varepsilon<<1$ measures the deviation from degeneracy, and the
limit $r_-\rightarrow r_+$ is obtained when $\varepsilon
\rightarrow 0$. Now, we introduce a new time coordinate $T$, $t=
T/(\varepsilon A)$, and a new radial coordinate $\chi$,
$r=\rho+\varepsilon \cosh\chi$, where $A\equiv
 A(\rho)=\rho^{-2}\left [ 1- \Lambda
 \left ( \rho^2+h \right )/3 \right ]>0$, with
$h\equiv h(\rho)$ defined in (\ref{aux function h(r)}). Then, in
the limit $\varepsilon \rightarrow 0$, from  (\ref{RN:metric
D-dim}) and (\ref{Fextreme D-dim}), we obtain the gravitational
field of the dS Bertotti-Robinson solution
\begin{eqnarray}
d s^2 = \frac{1}{A} \left (-\sinh^2\chi\, dT^2 +d\chi^2 \right ) +
 \frac{1}{B}\, d\Omega_{D-2}^2 \:.
 \label{D-dim:BertRob solution}
\end{eqnarray}
where $A$ and $B=1/\rho^2$ are related to $\Lambda$ and $Q$ by
\begin{eqnarray}
& & \Lambda = \frac{3}{(D-2)(D-1)}\left [-A+(D-3)^2 B \right ]\:,
 \nonumber \\
& & Q^2=\frac{(D-3)B+A}{(D-3)(D-2)B^{D-2}}\:. \label{D-dim:BertRob
solution:range AB}
\end{eqnarray}
The Maxwell field (\ref{RN:maxwell D-dim}) of the higher
dimensional dS Bertotti-Robinson solution is
 \begin{eqnarray}
 F=-Q_{\rm ADM}\,\frac{B^{(D-2)/2}}{A}\,\sinh \chi \,dT \wedge d\chi\:.
\label{D-dim:BertRob Maxwell}
\end{eqnarray}
So, if we give the parameters $\Lambda$, and $Q$ we can construct
the higher dimensional dS Bertotti-Robinson  solution, which  is
an exact solution of Einstein-Maxwell equations (\ref{D-dim:eqs of
motion Einstein}) with $\Lambda>0$ in $D$-dimensions. There is no
neutral ($Q=0$) Bertotti-Robinson solution.

Through a redefinition of coordinates, $\tau=\sqrt{A}\,T$ and
$\sinh^2 \chi=A\,R^2-1$, the spacetime (\ref{D-dim:BertRob
solution}) can be rewritten in new static coordinates as
\begin{eqnarray}
d s^2 = - (A\,R^2-1)\, dT^2 +\frac{dR^2}{A\,R^2-1}
 +\frac{1}{B}\, d\Omega_{D-2}^2,
 \label{D-dim:br2}
\end{eqnarray}
and the electromagnetic field changes also accordingly to the
coordinate transformation. Written in these coordinates, we
clearly see that the Bertotti-Robinson solution is the direct
topological product of $AdS_2 \times S^{D-2}$, i.e., of a
(1+1)-dimensional AdS spacetime with a ($D-2$)-sphere of fixed
radius $B^{-1/2}$.

\begin{figure}[h]
\includegraphics*[height=3cm]{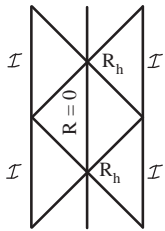}
   \caption{\label{Fig Ddim BR}
Carter-Penrose diagram of the Bertotti-Robinson solution in any
cosmological constant background. The zigzag line represents a
curvature singularity, ${\cal I}$ represents the infinity
($R=\infty$), $R_{\rm h}$ represents a horizon.
 }
\end{figure}

 The Carter-Penrose diagram of the $D$-dimensional
Bertotti-Robinson solution is sketched in Fig. \ref{Fig Ddim BR}.
In this diagram any point represents a ($D-2$)-sphere with fixed
radius $B^{-1/2}$. For a construction that starts with the
Carter-Penrose diagram of the dS black hole and leads to the
diagram of the Bertotti-Robinson solution see
\cite{OscLem_nariai}.

\vspace{0.5cm}

The higher dimensional flat Bertotti-Robinson solution is given by
the $\Lambda=0$ limit of the dS Bertotti-Robinson. It is described
by (\ref{D-dim:BertRob solution}) and (\ref{D-dim:BertRob
Maxwell}) with $A$ and $B$ being related to $Q$ by
\begin{eqnarray}
A=(D-3)^2\,Q^{-2/(D-3)},\quad {\rm and} \quad B = Q^{-2/(D-3)}.
\label{D-dim:BertRob L=0 solution:range AB}
\end{eqnarray}
Topologically this solution is also $AdS_2 \times S^{D-2}$, and is
an exact solution of Einstein-Maxwell equations
 (\ref{D-dim:eqs of motion Einstein}) with $\Lambda=0$ in $D$-dimensions. The
Carter-Penrose diagram of the higher dimensional flat
Bertotti-Robinson solution is also given by Fig. \ref{Fig Ddim
BR}.

%%%%%%%%%%%%%%%%%%%%%%%%%%%%%%%%%%%%%%%%%%%%%%%%%
\subsubsection{\label{sec:BH D-dim Nariai-Bertotti-Robinson}Higher
dimensional Nariai$-$Bertotti-Robinson solution}
%%%%%%%%%%%%%%%%%%%%%%%%%%%%%%%%%%%%%%%%%%%%%%%%%

In order to generate the higher dimensional
Nariai$-$Bertotti-Robinson solution from the near-ultracold black
hole we first go back to (\ref{Fextreme D-dim}) and rewrite it in
the form $f(r)=-P(r)(r-\rho)^2(r-\sigma)$, where $r=\rho$ is a
degenerate horizon of the black hole, $\sigma> \rho$ is the other
horizon, and $P(r)$ is a polynomial function of $r$. Then, we set
$\rho=\rho_{\rm u}-\varepsilon$ and $\sigma=\rho_{\rm u}+
\varepsilon$, with $\rho_{\rm u}$ defined in
 (\ref{def rho ultra}) and $\varepsilon<<1$ measuring the deviation
from degeneracy, and the limit $\rho \rightarrow \sigma$ is
obtained when $\varepsilon \rightarrow 0$. Now, we introduce a new
time coordinate $T$, $t= T/(2\varepsilon^2 P)$, and a new radial
coordinate $\chi$, $r=\rho_{\rm u}+\varepsilon \cos\left (
\sqrt{2\varepsilon P}\,\chi \right )$, where
 $P\equiv P(\rho_{\rm u})>0$.
Then, taking the limit $\varepsilon \rightarrow 0$ of
(\ref{RN:metric D-dim}), we obtain the gravitational field of the
Nariai$-$Bertotti-Robinson solution, $d s^2 =-\chi^2\, dT^2
+d\chi^2 +\rho_{\rm u}^{\,2}\, d\Omega_{D-2}^2$, where $\chi$ runs
from $0$ to $+\infty$, and the Maxwell field
 (\ref{RN:maxwell D-dim}) of the higher dimensional Nariai$-$Bertotti-Robinson
solution is $F=Q_{\rm ADM}\,\rho_{\rm u}^{-D+2}\,\chi \,dT \wedge
d\chi$. Now, the spacetime factor $-\chi^2\, dT^2 +d\chi^2$ is
just ${\mathbb{M}}^{1,1}$ (2-dimensional Minkowski spacetime) in
Rindler coordinates. Therefore, under the usual coordinate
transformation $\chi=\sqrt{x^2-t^2}$ and $T={\rm arctanh}(t/x)$,
we can write the higher dimensional Nariai$-$Bertotti-Robinson
solution in its simplest form,
\begin{eqnarray}
d s^2 =-dt^2 +dx^2 +\rho_{\rm u}^{\,2}\, d\Omega_{D-2}^2 \:,
 \label{D-dim:Nariai BertRob solution}
\end{eqnarray}
where $\rho_{\rm u}$ is defined in (\ref{def rho ultra}), and
\begin{eqnarray}
 F=-\frac{Q_{\rm ADM}}{\rho_{\rm u}^{D-2}} \,dt \wedge
 dx\:,
\label{D-dim:Nariai BertRob Maxwell}
\end{eqnarray}
where $Q_{\rm ADM}$ is given by (\ref{ADM hairs D-dim}) and
(\ref{UltracoldDdim:range}). So, if we give  $\Lambda$ we can
construct the higher dimensional Nariai$-$Bertotti-Robinson
solution.  This solution is the direct topological product of
${\mathbb{M}}^{1,1}\times S^{D-2}$, and is an exact solution of
Einstein-Maxwell equations (\ref{D-dim:eqs of motion Einstein})
with $\Lambda>0$ in $D$-dimensions. Its causal diagram is equal to
the causal diagram of the Rindler solution. This solution belongs
to the class of solutions discussed in detail in \cite{PlebHac}
(for $D=4$), and thus it can very appropriately be called a
Pleba\'nski-Hacyan solution \cite{OrtaggioPodolsky}.

%%%%%%%%%%%%%%%%%%%%%%%%%%%%%%%%%%%%%%%%%%%%%%%%%
\section{\label{sec:Solutions D-dim cosmolog AdS}Higher dimensional
extreme A\lowercase{d}S black holes and anti-Nariai like
solutions}
%%%%%%%%%%%%%%%%%%%%%%%%%%%%%%%%%%%%%%%%%%%%%%%%%
In order to generate the higher dimensional anti-Nariai, and the
two AdS$-$Bertotti-Robinson solutions one needs first to carefully
find the values of the mass and of the charge for which one has
extreme AdS black holes. We will do this in Sec.
 \ref{sec:BH D-dim cosmolog AdS}, and in Sec. \ref{sec:BH D-dim extremal limits AdS}
we will generate the anti-Nariai like solutions from the extremal
limits of the near-extreme black holes.
%%%%%%%%%%%%%%%%%%%%%%%%%%%%%%%%%%%%%%%%%%%%%%%%%
\subsection{\label{sec:BH D-dim cosmolog AdS}Higher dimensional
extreme black holes in an asymptotically A\lowercase{d}S
background}
%%%%%%%%%%%%%%%%%%%%%%%%%%%%%%%%%%%%%%%%%%%%%%%%%

In a higher dimensional asymptotically anti-de Sitter background,
$\Lambda< 0$, the Einstein-Maxwell equations
 (\ref{D-dim:eqs of motion Einstein}) allow a three-family of static black hole
solutions, parameterized by the constant $k$ which can take the
values $1,0,-1$, and whose gravitational field is described by
\begin{eqnarray}
d s^2 = - f(r)\, dt^2 +f(r)^{-1}\,dr^2+r^2 (d\,\Omega_{D-2}^k)^2,
 \label{AdS bH D-dim}
\end{eqnarray}
where
\begin{eqnarray}
f(r) =
k-\frac{\Lambda}{3}\,r^2-\frac{M}{r^{D-3}}+\frac{Q^2}{r^{2(D-3)}}\:,
 \label{RN:f AdS D-dim}
\end{eqnarray}
and for $k=1$, $k=0$ and $k=-1$ one has, respectively,
\begin{eqnarray}
(d\Omega_{D-2}^k)^2 \!\!\!&=& \!\!\!
d\theta_1^2+\sin^2\theta_1\,d\theta_2^2+ \cdots
+\prod_{i=1}^{D-3}\sin^2\theta_i\,d\theta_{D-2}^2\,,  \nonumber \\
(d\Omega_{D-2}^k)^2 \!\!\!&=& \!\!\! d\theta_1^2+d\theta_2^2+
d\theta_3^2+\cdots +d\theta_{D-2}^2\,,        \nonumber \\
(d\Omega_{D-2}^k)^2 \!\!\!&=& \!\!\!
d\theta_1^2+\sinh^2\theta_1\,d\theta_2^2+\! \cdots \!
+\!\!\prod_{i=1}^{D-3}\sinh^2\theta_i\,d\theta_{D-2}^2\,. \nonumber \\
& &
 \label{angular AdS bh D-dim}
\end{eqnarray}
Thus, the family with $k=1$ yields AdS black holes with spherical
topology found in \cite{tangherlini}. The family with $k=0$ yields
AdS black holes with planar, cylindrical  or toroidal (with genus
$g\geq 1$) topology that are the higher dimensional counterparts
(introduced in \cite{Birmingham-TopDdim,Toroidal_Q_Ddim}) of the
4-dimensional black holes found and analyzed in
\cite{Toroidal_4D,Toroidal_elect_4D}. Finally, the family with
$k=-1$ yields AdS black holes with hyperbolic, or toroidal
topology with genus $g\geq 2$ that are the higher dimensional
counterparts (introduced in
\cite{Birmingham-TopDdim,DehghaniRotTopol} in the neutral case) of
the 4-dimensional black holes analyzed in \cite{topological}. The
solutions with non-spherical topology (i.e., with $k=0$ and
$k=-1$) do not have counterparts in a $\Lambda=0$ or in a
$\Lambda>0$ background.

The mass parameter $M$ and the charge parameter $Q$ are related to
the ADM hairs, $M_{\rm ADM}$ and $Q_{\rm ADM}$, by (\ref{ADM hairs
D-dim}).  These black holes can have at most two horizons.
Following a similar procedure as the one sketched in Sec.
{\ref{sec:BH D-dim dS}, we find the mass parameter $M$ and the
charge parameter $Q$ of the extreme black holes as a function of
the degenerate horizon at $r=\rho$
\begin{eqnarray}
& & M=2\rho^{D-3} \left ( k-\frac{D-2}{D-3}\,\frac{\Lambda}{3}
\rho^2
\right)\:,  \nonumber \\
& & Q^2 =\rho^{2(D-3)} \left (
k-\frac{D-1}{D-3}\,\frac{\Lambda}{3} \rho^2 \right)\:.
 \label{mq AdS D-dim}
 \end{eqnarray}
This equation is not the main result of Sec.
 \ref{sec:Solutions D-dim cosmolog AdS}. However, it constitutes a new result that we had to
find in our way into the generation of the anti-Nariai like
solutions. For an alternative and complementary description of the
extreme AdS black holes in $D$-dimensions see
\cite{KodamaIshibashi}. For $D=5$, and only in this case, we were
able to write $f(r)$ in the extreme case as a function of the
degenerate horizon $\rho$. We write this expression here since it
might be useful for future work,
\begin{eqnarray}
f(r)=-\frac{\Lambda}{3}\frac{1}{r^4}(r-\rho)^2(r+\rho)^2
 \left ( r^2+2\rho^2-k\frac{3}{\Lambda}\right ).
 \label{f(r) 5D AdS D-dim}
 \end{eqnarray}
For $D\geq 6$ we were not able to to write $f(r)$  as a function
of $\rho$ since one has to deal with polynomial functions with
degree higher than four.

%%%%%%%%%%%%%%%%%%%%%%%%%%%%%%%%%%%%%%%%%%%%%%%%%
\subsubsection{\label{sec:BH D-dim AdS spherical}Higher dimensional
AdS black holes with spherical topology}
%%%%%%%%%%%%%%%%%%%%%%%%%%%%%%%%%%%%%%%%%%%%%%%%%

When $k=1$, one has $0<\rho<+\infty$ and $M$ and $Q$ in
 (\ref{mq AdS D-dim}) are positive parameters. The ranges of $M$ and $Q$
that represent extreme and nonextreme black holes are sketched in
Fig. \ref{range mq AdS spher+toroid bh D-dim}.
\begin{figure}[h]
\centering
\includegraphics*[height=4cm]{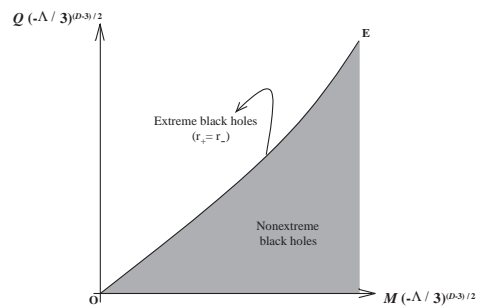}
   \caption{\label{range mq AdS spher+toroid bh D-dim}
Range of $M$ and $Q$ for which one has a nonextreme black hole
(shaded region), and an extreme black hole with $r_+ = r_-$ (line
$OE$) in the AdS case with spherical topology ($k=1$) or with
planar, cylindrical or toroidal topology ($k=0$). The non-shaded
region represents naked singularities.
 }
\end{figure}

The Carter-Penrose diagram of the nonextreme
AdS$-$Reissner-Nordstr\"{o}m black hole is sketched in Fig.
\ref{Fig spherical Q AdS}.(a), and the one of the extreme
AdS$-$Reissner-Nordstr\"{o}m black hole is represented in Fig.
\ref{Fig spherical Q AdS}.(b). The Carter-Penrose diagram of the
AdS-Schwarzschild black hole is drawn in Fig.
 \ref{Fig spherical Q=0 AdS}. In these
diagrams each point represents a ($D-2$)-sphere of radius $r$.
\begin{figure}[h]
\includegraphics*[height=6cm]{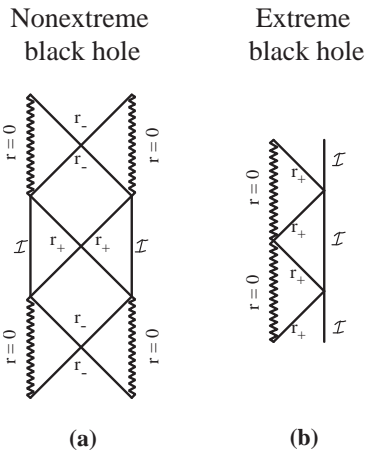}
   \caption{\label{Fig spherical Q AdS}
Carter-Penrose diagrams of the charged ($Q\neq 0$) AdS black holes
with $k=1$ and $k=0$. The zigzag line represents a curvature
singularity, ${\cal I}$ represents the infinity ($r=\infty$),
$r_+$ represents a black hole event horizon, and $r_-$ represents
a Cauchy horizon.
 }
\end{figure}
\begin{figure}[h]
\includegraphics*[height=2cm]{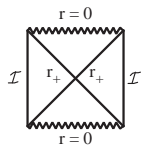}
   \caption{\label{Fig spherical Q=0 AdS}
Carter-Penrose diagrams of the neutral ($Q= 0$) AdS  black holes
with $k=1$ and $k=0$. The zigzag line represents a curvature
singularity, ${\cal I}$ represents the infinity ($r=\infty$), and
$r_+$ represents a black hole event horizon.
 }
\end{figure}
%%%%%%%%%%%%%%%%%%%%%%%%%%%%%%%%%%%%%%%%%%%%%%%%%
\subsubsection{\label{sec:BH D-dim AdS toroidal}Higher dimensional
AdS black holes  with toroidal, cylindrical or planar topology}
%%%%%%%%%%%%%%%%%%%%%%%%%%%%%%%%%%%%%%%%%%%%%%%%%

When $k=0$, one has $0<\rho<+\infty$, and $M$ and $Q$ in
 (\ref{mq AdS D-dim}) are positive parameters. The ranges of $M$ and $Q$
that represent extreme and nonextreme black holes are sketched in
Fig. \ref{range mq AdS spher+toroid bh D-dim}.

 The Carter-Penrose diagram of the
nonextreme charged AdS black hole with $k=0$ is sketched in Fig.
\ref{Fig spherical Q AdS}.(a), and the one of the extreme charged
AdS black hole  with $k=0$ is represented in Fig. \ref{Fig
spherical Q AdS}.(b). The Carter-Penrose diagram of the neutral
AdS black hole with $k=0$ is drawn in Fig. \ref{Fig spherical Q=0
AdS}. In these diagrams each point represents a ($D-2$)-plane, or
a ($D-2$)-cylinder or a ($D-2$)-torus.
%%%%%%%%%%%%%%%%%%%%%%%%%%%%%%%%%%%%%%%%%%%%%%%%%
\subsubsection{\label{sec:BH D-dim AdS topological}Higher dimensional
AdS black holes with hyperbolic topology}
%%%%%%%%%%%%%%%%%%%%%%%%%%%%%%%%%%%%%%%%%%%%%%%%%

When $k=-1$, the condition that $Q^2\geq 0$ demands that
 $\rho_{\rm min}\leq \rho<+\infty$, where
\begin{eqnarray}
\rho_{\rm min}=\sqrt{-\frac{D-3}{D-1}\frac{3}{\Lambda}}.
 \label{Rmin D-dim}
 \end{eqnarray}
For  $\rho=\rho_{\rm min}$, the extreme black hole has no electric
charge ($Q=0$) and its mass is negative, $M=-4\rho_{\rm
min}^{D-3}/(D-1)$. For $\rho=\rho_0$, where
\begin{eqnarray}
\rho_0=\sqrt{-\frac{D-3}{D-2}\frac{3}{\Lambda}}\,,
 \label{R0 D-dim}
 \end{eqnarray}
the extreme black hole has no mass ($M=0$) and its charge is given
by $Q=\rho_0^{D-3}/\sqrt{D-2}$. The ranges of $M$ and $Q$ that
represent extreme and nonextreme black holes are sketched in Fig.
\ref{range mq AdS hyp bh D-dim}.
\begin{figure}[h]
\centering
\includegraphics*[height=4cm]{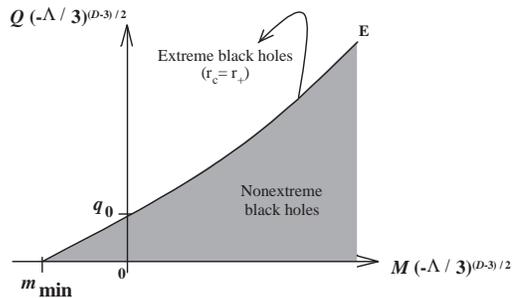}
   \caption{\label{range mq AdS hyp bh D-dim}
Range of $M$ and $Q$ for which one has a nonextreme black hole,
and an extreme black hole with $r_{\rm c} = r_+$ in the AdS case
with hyperbolic topology ($k=-1$). The non-shaded region
represents a naked singularity. One has $m_{\rm
min}=-\frac{4}{D-1}(\frac{D-3}{D-2})^{(D-3)/2}$, and
$q_0=(\frac{D-3}{D-2})^{(D-3)/2}/\sqrt{D-2}$.
 }
\end{figure}

In what concerns the causal structure of these solutions, when
$M=0$ and $Q=0$, the solution has an horizon that we identify as a
cosmological horizon ($r_{\rm c}$) since it is present when the
mass and charge vanish. In this case $r=0$ is not a curvature
singularity, but can be regarded as a topological singularity (see
Brill, Louko, and Peldan in \cite{topological} for a detailed
discussion). The Carter-Penrose diagram of this solution is drawn
in Fig. \ref{Fig hyp Q=0 AdS}, as long as we interpret the zigzag
line as being a topological singularity. When $Q=0$ and $M>0$, the
solution still has a single horizon, the same cosmological horizon
that is present in the latter case. However, now a curvature
singularity is present at $r=0$.  The corresponding Carter-Penrose
diagram of this solution is represented in Fig. \ref{Fig hyp Q=0
AdS}. The most interesting $Q=0$ solutions are present when their
mass is negative. In this case one can have a black hole solution
with a black hole horizon and a cosmological horizon [see Fig.
\ref{Fig hyp Q AdS}.(a)], or an extreme black hole, in which the
two above horizons merge [see Fig. \ref{Fig hyp Q AdS}.(b)]. When
$Q\neq 0$, one can have a black hole solution with a black hole
horizon and a cosmological horizon (see Fig. \ref{Fig hyp Q
AdS}.(a) for the corresponding causal diagram), or an extreme
black hole, in which the two above horizons merge (see Fig.
\ref{Fig hyp Q AdS}.(b) for the corresponding causal diagram).
Note that the presence of the charge does not introduce an extra
horizon, contrary to what usually  occurs in the other black hole
solutions.
\begin{figure}[h]
\includegraphics*[height=2cm]{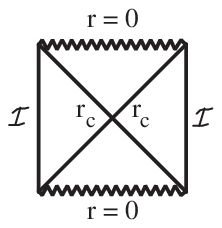}
   \caption{\label{Fig hyp Q=0 AdS}
Carter-Penrose diagrams of the neutral AdS  black holes with
$k=-1$. The zigzag line represents a curvature singularity, ${\cal
I}$ represents the infinity ($r=\infty$), $r_+$ represents a black
hole event horizon, and $r_-$ represents a Cauchy horizon.
 }
\end{figure}
\begin{figure}[h]
\includegraphics*[height=6cm]{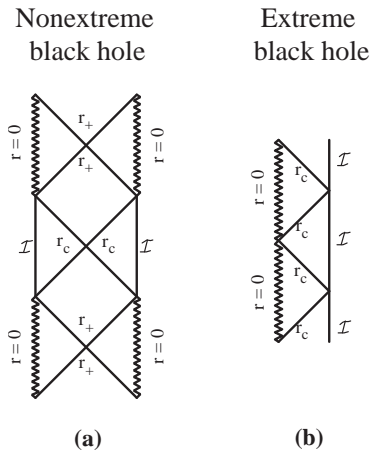}
   \caption{\label{Fig hyp Q AdS}
Carter-Penrose diagrams of the charged AdS black holes with
$k=-1$. The zigzag line represents a curvature singularity, ${\cal
I}$ represents the infinity ($r=\infty$), $r_+$ represents a black
hole event horizon, and $r_-$ represents a Cauchy horizon.
 }
\end{figure}
%%%%%%%%%%%%%%%%%%%%%%%%%%%%%%%%%%%%%%%%%%%%%%%%%
\subsection{\label{sec:BH D-dim extremal limits AdS}Extremal
limits of the higher dimensional A\lowercase{d}S black holes}
%%%%%%%%%%%%%%%%%%%%%%%%%%%%%%%%%%%%%%%%%%%%%%%%%

In this subsection, we will consider the extremal limits of the
near-extreme higher dimensional AdS black holes. This procedure
leads to the generation of the higher dimensional anti-Nariai
solution and to the higher dimensional AdS Bertotti-Robinson
solutions.
 To achieve our aim, we first go back to the extreme case of
(\ref{RN:f AdS D-dim}) and rewrite it in the form
$f(r)=A(r)(r-\rho)^2$, where $r=\rho$ is the degenerate horizon of
the black hole, and $A(r)$ is a polynomial function of $r$. Then,
we introduce a new time coordinate $T$, $t= T/(\varepsilon A)$,
and a new radial coordinate $\chi$, $r=\rho+\varepsilon
\cosh\chi$, where $A\equiv A(\rho)$, and $\varepsilon<<1$ measures
the deviation from degeneracy. Finally, taking the limit
$\varepsilon \rightarrow 0$ in (\ref{AdS bH D-dim}) yields the
gravitational field of the new higher dimensional solutions,
\begin{eqnarray}
d s^2 = \frac{1}{A} \left (-\sinh^2\chi\, dT^2 +d\chi^2 \right ) +
 \frac{1}{B}\, (d\Omega^{k}_{D-2})^2 \:,
 \label{D-dim:AdS NBR solution}
\end{eqnarray}
where $k=1,0,-1$ in the spherical, cylindrical and hyperbolic
cases, respectively, and $A$ and $B$ are constants related to
$\Lambda$ and $Q$ by
\begin{eqnarray}
& &  \Lambda =-\frac{3}{(D-2)(D-1)}\left [ A-k\,(D-3)^2 B \right
]\:,
 \nonumber \\
& & Q^2=\frac{A+k\,(D-3)B}{(D-3)(D-2)B^{D-2}}\:,
\label{D-dim:range AB NBR AdS}
\end{eqnarray}
In the new coordinate system, the Maxwell field
 (\ref{RN:maxwell D-dim}) of the solutions is
 \begin{eqnarray}
 F=-Q_{\rm ADM}\,\frac{B^{(D-2)/2}}{A}\,\sinh \chi \,dT \wedge d\chi\:.
\label{D-dim:NBR AdS Maxwell}
\end{eqnarray}
Equations
 (\ref{D-dim:AdS NBR solution})-(\ref{D-dim:NBR AdS Maxwell})
describe three exact solutions of the Einstein-Maxwell equations
(\ref{D-dim:eqs of motion Einstein}) with $\Lambda<0$ in
$D$-dimensions, that we discuss in the following subsections.

%%%%%%%%%%%%%%%%%%%%%%%%%%%%%%%%%%%%%%%%%%%%%%%%%
\subsubsection{\label{sec:BR D-dim AdS spherical}Higher dimensional
AdS Bertotti-Robinson solution with spherical topology}
%%%%%%%%%%%%%%%%%%%%%%%%%%%%%%%%%%%%%%%%%%%%%%%%%

The $k=+1$ case describes the AdS Bertotti-Robinson solution with
spherical topology. This solution is the direct topological
product of $AdS_2 \times S^{D-2}$, i.e., of a (1+1)-dimensional
AdS spacetime with a ($D-2$)-sphere of fixed radius $B^{-1/2}$.
The Carter-Penrose diagram of this $D$-dimensional spherical AdS
Bertotti-Robinson solution is sketched in Fig. \ref{Fig Ddim BR}.
In this diagram any point represents a ($D-2$)-sphere with fixed
radius $B^{-1/2}$.

%%%%%%%%%%%%%%%%%%%%%%%%%%%%%%%%%%%%%%%%%%%%%%%%%
\subsubsection{\label{sec:BR D-dim AdS toroidal}Higher dimensional
AdS Bertotti-Robinson solution with toroidal, cylindrical or
planar topology}
%%%%%%%%%%%%%%%%%%%%%%%%%%%%%%%%%%%%%%%%%%%%%%%%%

The $k=0$ case describes the AdS Bertotti-Robinson solution with
toroidal, cylindrical or planar topology, also known as
Pleba\'nski-Hacyan solution \cite{PlebHac}. This solution is the
direct topological product of $AdS_2 \times \mathbb{E}^{D-2}$,
i.e., of a (1+1)-dimensional AdS spacetime with a ($D-2$)
Euclidean space. The Carter-Penrose diagram of this
$D$-dimensional toroidal Bertotti-Robinson solution is sketched in
Fig. \ref{Fig Ddim BR}. In this diagram any point represents a
($D-2$)-plane, a ($D-2$)-cylinder, or a ($D-2$)-torus with fixed
size.

%%%%%%%%%%%%%%%%%%%%%%%%%%%%%%%%%%%%%%%%%%%%%%%%%
\subsubsection{\label{sec:anti-Nariai D-dim}Higher dimensional
anti-Nariai solution}
%%%%%%%%%%%%%%%%%%%%%%%%%%%%%%%%%%%%%%%%%%%%%%%%%

Finally, the $k=-1$ case describes the the higher dimensional
anti-Nariai solution. This solution is the direct topological
product of $AdS_2 \times H^{D-2}$, i.e., of a (1+1)-dimensional
AdS spacetime with a ($D-2$)-hyperboloid with a fixed size,
$B^{-1/2}$. The $k=-1$ case is the only one that admits a solution
with $Q=0$. This neutral anti-Nariai solution satisfies
$A=-\Lambda (D-1)/3$ and $B=-\Lambda (D-1)/(3D-9)$. The
$\Lambda=0$ limit of the anti-Nariai solution is $D$-dimensional
Minkowski spacetime as occurs with the $D=4$ solution (see
\cite{OscLem_nariai}). The Carter-Penrose diagram of this
$D$-dimensional anti-Nariai solution is sketched in Fig.
 \ref{Fig Ddim BR}. In this diagram any point represents a
($D-2$)-hyperboloid with fixed size, $B^{-1/2}$.

%%%%%%%%%%%%%%%%%%%%%%%%%%%%%%%%%%%%%%%%%%%%%%%%%%%%%%%%%%
%%%%%%%%%%%%%%%%%%%%%%%%%%%%%
\section{\label{sec:Conc}Conclusion}
%%%%%%%%%%%%%%%%%%%%%%%%%%%%%
%%%%%%%%%%%%%%%%%%%%%%%%%%%%%%%%%%%%%%%%%%%%%%%%%%%%%%%%%%

We have constructed all the higher dimensional solutions that are
the topological product of two manifolds of constant curvature,
and that can be generated from the extremal limits of the
near-extreme black holes. Our analysis yields explicit results
that apply to any dimension $D\geq 4$. These solutions include the
$D$-dimensional counterparts of the well-known Nariai,
Bertotti-Robinson, anti-Nariai, and Pleba\'nski-Hacyan solutions.
In order to achieve our aim we had to find the values of the mass
and of the charge for which one has extreme black holes in an
asymptotically de Sitter (dS) and in an asymptotically anti-de
Sitter (AdS) higher dimensional spacetime. This is not an easy
task in a $D$-dimensional background since one has to deal with
polynomial functions with degree higher than four.

Nowadays, one of the main motivations to study higher dimensional
asymptotically AdS or dS black holes is related with the AdS/CFT
and dS/CFT correspondences.  In particular, the higher dimensional
cosmological black holes are useful to study the dynamics of
Friedmann-Robertson-Walker branes in the framework of (A)dS/CFT
correspondence (for a review see, e.g., \cite{NojiriOdintsov}).
The solutions discussed in this paper, being extremal limits of
higher dimensional black holes, might also be interesting in this
context. They might also be useful for the discussion of modified
gravities. Another research area that might follow from this paper
is the study of impulsive waves in the background of the
spacetimes that we presented, in a direct generalization of the
analysis done for four dimensions in \cite{OrtaggioPodolsky}. As a
last example of application of these higher dimensional solutions,
we mention the study of their classical stability, i.e., the exact
analytical analysis of their quasinormal modes. Finally, the
quantum stability of these Nariai-like solutions will be discussed
in \cite{oscarvitorjosePC}.

%%%%%%%%%%%%%%%%%%%%%%%%%%%%%%%%%%%%%%%%%%%%%%%%%

%%%%%%%%%%%%%%%%%%%%%%%%%%%%%%%%%%%%%%%%%%%%%%%%%
\begin{acknowledgments}

The authors would like to thank Marcello Ortaggio for a critical
reading of the manuscript and for pointing out their attention to
references \cite{PlebHac,FickenCahenDefrise}. This work was
partially funded by Funda\c c\~ao para a Ci\^encia e Tecnologia
(FCT) through project CERN/FIS/43797/2001 and PESO/PRO/2000/4014.
VC and OJCD also acknowledge finantial support from the FCT
through PRAXIS XXI programme. JPSL thanks Observat\'orio Nacional
do Rio de Janeiro for hospitality.

\end{acknowledgments}

%%%%%%%%%%%%%%%%%%%%%%%%%%%%%%%%%%%%%%%%%%%%%%%%%%%%%%%%%%
%%%%%%%%%%%%%%%%%%%%%%%%%%%%%
%\appendix*
%\section{\label{sec:Sub-Maximal zeros}Properties of the general nonextreme solution}
%%%%%%%%%%%%%%%%%%%%%%%%%%%%%

%%%%%%%%%%%%%%%%%%%%%%%%%%%%%%%%%%%%%%%%%%%%%%%%%
%%%%%%%%%%%%%%%%%%%%%%%%%%%%%%%%%%%%%%%%%%%%%%%%%


\begin{thebibliography}{100}

\bibitem{hamed} N. Arkani-Hamed, S. Dimopoulos and G. Dvali,
Phys. Lett. {\bf B429}, 263 (1998); Phys. Rev. D{\bf 59}, 086004
(1999);

I. Antoniadis, N. Arkani-Hamed, S. Dimopoulos and G. Dvali, Phys.
Lett. {\bf B436}, 257 (1998).

\bibitem{bhprod} P. C. Argyres, S. Dimopoulos and J. March-Russell
Phys. Lett. {\bf B441}, 96 (1998); S. Dimopoulos, G. Landsberg,
Phys. Rev. Lett. {\bf 87}, 161602 (2001); Neutrino cosmic rays
could also provide observable examples of black hole production in
high-energy particle collisions. See for example, J. L. Feng, A.
D. Shapere, Phys. Rev. Lett. {\bf 88}, 021303 (2002).

\bibitem{tails} A. O. Barvinsky, S. N. Solodukhin,
Nucl. Phys. B{\bf 675}, 159 (2003); V. Cardoso, S. Yoshida, O. J.
C. Dias, J. P. S. Lemos, Phys. Rev. D {\bf 68}, R061503 (2003). V.
Cardoso, O. J. C. Dias and J. P. S. Lemos, Phys. Rev. D {\bf 67},
064026 (2003).


\bibitem{tangherlini} F. R. Tangherlini,
Nuovo Cim. {\bf 27}, 636 (1963).

\bibitem{myersMP} R. C. Myers, Phys. Rev. D {\bf 35}, 455 (1987).
\bibitem{lemoszanchinMP} J. P. S. Lemos, V. T. Zanchin, {\it Higher dimensional
Majumdar-Papapetrou stars}, in preparation (2004).
\bibitem{myersperry} R. C. Myers and M. J. Perry,
Annals Phys. {\bf 172}, 304 (1986).
\bibitem{EmparanMyers} R. Emparan, R. C. Myers, JHEP {\bf 0309} 025 (2003).
\bibitem{IdaUchida} D. Ida, Y. Uchida, Phys. Rev. D {\bf 68},
104014 (2003).

\bibitem{Toroidal_4D}  J. P. S. Lemos, Class. Quantum Grav. {\bf 12}, 1081 (1995);
Phys. Lett. B{\bf 353}, 46 (1995); P. M. S\'a, A. Kleber, J. P. S.
Lemos, Class. Quantum Grav. {\bf 13}, 125 (1996);  P. M. S\'a, J.
P. S. Lemos, Phys. Lett. B{\bf 423}, 49 (1998).
\bibitem{Toroidal_elect_4D} J. P. S. Lemos, V. T. Zanchin,  Phys. Rev. D{\bf 54}, 3840
(1996); O. J. C. Dias, J. P. S. Lemos, Phys. Rev. D{\bf 64},
064001 (2001).
\bibitem{Toroidal_mag_4D}  O. J. C. Dias, J. P. S. Lemos, Class. Quantum Grav. {\bf 19}, 2265 (2002);
 Phys. Rev. D{\bf 66}, 024034 (2002).
\bibitem{KMVanz} D. Klemm, V. Moretti, L. Vanzo, Phys. Rev.
D{\bf 57}, 6127 (1998).

\bibitem{topological} S. \AA minnenborg, I. Bengtsson, S. Holst,
P. Peld\'an,  Class. Quantum Grav. {\bf 13}, 2707 (1996); D. R.
Brill, Helv. Phys. Acta {\bf 69}, 249 (1996); S. L. Vanzo, Phys.
Rev. D {\bf 56}, 6475 (1997); D. R. Brill, J. Louko, P. Peld\'an,
Phys. Rev. D {\bf 56}, 3600 (1997); R. B. Mann, Class. Quantum
Grav. {\bf 14}, 2927 (1997); S. Holst, P. Peldan, Class. Quantum
Grav. {\bf 14}, 3433 (1997).


\bibitem{Birmingham-TopDdim} D. Birmingham,
Class. Quantum Grav. {\bf 16}, 1197 (1999); R. G. Cai, K. S. Soh,
Phys. Rev. D{\bf 59}, 044013 (1999).
\bibitem{Toroidal_Q_Ddim} A. Chamblin, R. Emparan, C. V. Johnson,
R. C. Myers,  Phys. Rev. D {\bf 60}, 064018 (1999); Phys. Rev. D
{\bf 60}, 104026 (1999);  A. M.  Awad, Class. Quantum Grav. {\bf
20}, 2827 (2003).
\bibitem{Dehghani-Toroidal_magJ_Ddim} M. H. Dehghani,
Phys. Rev. D {\bf 69}, 044024 (2004).
\bibitem{DehghaniRotTopol} M. H. Dehghani, Phys. Rev. D {\bf 65},
124002 (2002); Phys. Rev. D {\bf 65}, 104003 (2002).

\bibitem{KodamaIshibashi} H. Kodama, A. Ishibashi,
Prog. Theor. Phys. {\bf 111}, 29 (2004).

\bibitem{Nariai} H. Nariai, Sci. Rep. Tohoku Univ. {\bf 34}, 160 (1950);
Sci. Rep. Tohoku Univ. {\bf 35}, 62 (1951).

\bibitem{BertRob} B. Bertotti, Phys. Rev. {\bf 116}, 1331
(1959); I. Robinson, Bull. Acad. Polon. {\bf 7}, 351 (1959).

\bibitem{CaldVanZer} M. Caldarelli, L. Vanzo, Z. Zerbini,
{\it The extremal limit of D-dimensional black holes}, in {\it
Geometrical Aspects of Quantum Fields}, edited by A. A. Bytsenko,
A. E. Gon\c calves, B. M. Pimentel (World Scientific, Singapore,
2001); {\tt hep-th/0008136}; N. Dadhich, {\it On product spacetime
with 2-sphere of constant curvature}, {\tt gr-qc/0003026}.
\bibitem{PlebHac} J. F. Pleba\'nski, S. Hacyan, J. Math. Phys. {\bf 20},
1004 (1979).

\bibitem{Bousso60y} S. Nojiri, S. D. Odintsov, Int. J.
Mod. Phys. A {\bf 14}, 1293 (1999); Phys. Rev. D {\bf 59}, 044026
(1999); R. Bousso, Phys. Rev. D {\bf 58}, 083511 (1998); {\it
Adventures in de Sitter space}, {\tt hep-th/0205177}.
\bibitem{OrtaggioPodolsky} M. Ortaggio, Phys. Rev. D {\bf 65}, 084046 (2002);
M. Ortaggio, J. Podolsk\'y, Class. Quantum Grav. {\bf 19} 5221
(2002); Class. Quantum Grav. {\bf 20} 1685 (2003).
\bibitem{OscLem_nariai}  O. J. C. Dias, J. P.
S. Lemos, Phys. Rev. D {\bf 68}, 061503 (2003).

\bibitem{FickenCahenDefrise} F. A. Ficken, Ann. Math. {\bf 40},
892 (1939); M. Cahen, L. Defrise, Commun. Math. Phys. {\bf 11}, 56
(1968).

\bibitem{GinsPerry} P. Ginsparg, M. J. Perry, Nucl. Phys. B {\bf 222}, 245
(1983).
\bibitem{Nariai_Q}
S. W. Hawking, S. F. Ross,  Phys. Rev. D {\bf 52}, 5865 (1995); R.
B. Mann, S. F. Ross, Phys. Rev. D {\bf 52}, 2254 (1995).


  \bibitem{Rom}L. J. Romans, Nucl. Phys. B {\bf
383}, 395 (1992).

\bibitem{GibHawdSbh} G. Gibbons, S. Hawking, Phys. Rev. D {\bf 15},
2738 (1977).
\bibitem{Lake} K. Lake, R. Roeder, Phys. Rev. D {\bf 15}, 3513 (1977).

\bibitem{NojiriOdintsov} S. Nojiri, S. D. Odintsov, JHEP {\bf 0112}
033 (2001); S. Nojiri, S. D. Odintsov, S. Ogushi, Phys. Rev. D
{\bf 65}, 023521 (2002); Int. J. Mod. Phys. A{\bf 17}, 4809
(2002).

\bibitem{oscarvitorjosePC} V. Cardoso, O. J. C. Dias, J. P.
S. Lemos,  {\it Pair creation of higher dimensional de Sitter
black holes}, to be submitted (2004).






\end{thebibliography}
\end{document}